\documentclass{article}
\usepackage{spconf,amsmath,amssymb,graphicx,booktabs,placeins,needspace,float}
\usepackage[hypertexnames=false,hidelinks]{hyperref}
\ninept
\let\originalthebibliography\thebibliography
\renewcommand{\thebibliography}[1]{\originalthebibliography{#1}\interlinepenalty=10000\setlength{\itemsep}{-1.75pt}\setlength{\parsep}{0pt}}

\title{SHINE: Sequential Hierarchical Integration Network for EEG and MEG}
\name{\shortstack{Xiran Xu$^{1,3}$ \quad Yujie Yan$^{2,3}$ \quad Songyi Li$^{1,3}$ \quad Linze Zheng$^{1,3}$ \\
Zifeng Zhang$^{2,3}$ \quad Mochu Dong$^{1,3}$ \quad Jing Chen$^{1,2,3,*}$}}
\address{$^{1}$Speech and Hearing Research Center, School of Intelligence Science and Technology, Peking University\\
$^{2}$Center for BioMed-X Research, Academy for Advanced Interdisciplinary Studies, Peking University\\
$^{3}$National Key Laboratory of General Artificial Intelligence\\
$^{*}$ Corresponding author. E-mail address: chenj@cis.pku.edu.cn}

\begin{document}
\setlength{\tabcolsep}{0pt}
\maketitle
\setlength{\tabcolsep}{3.0pt}

\begin{abstract}
How natural speech is represented in the brain constitutes a major challenge for cognitive neuroscience. Reconstructing the speech envelope and Mel spectrogram from EEG and MEG provides a time-resolved way to study its temporal and spectral structure. Speech-related neural activity spans sensors and temporal scales; extracting these representations while adapting the use of context to each acoustic target is a central problem in speech reconstruction. We propose SHINE, a \textbf{S}equential \textbf{H}ierarchical \textbf{I}ntegration \textbf{N}etwork for \textbf{E}EG and MEG. A residual sensor adapter unifies input dimensions, intermediate dilated-block states retain temporal depth, and a target- and time-dependent gate fuses local hierarchical and attention-enhanced context predictions. Across two EEG and two MEG datasets, SHINE has the highest mean envelope and mean-Mel Pearson correlations among nine local baseline implementations on all eight dataset--metric combinations. SHINE also placed \textbf{second} in the speech-detection Extended Track of the \textbf{NeurIPS 2025 PNPL Competition}. Code will be released at \url{https://github.com/xuxiran/SHINE}.
\end{abstract}


\begin{figure*}[!t]
\centering
\includegraphics[width=0.98\textwidth]{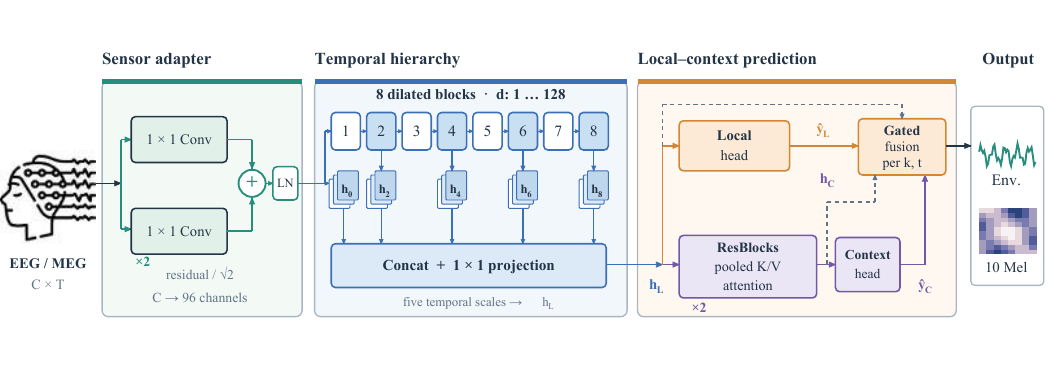}
\caption{SHINE architecture. The adapter's lower path has two $1\!\times\!1$ convolutions with GELU between them (``$\times2$'' in the diagram). Its output $h_0$ and four intermediate dilated-block states form five temporal depths that project to local features $h_L$; two pooled-attention blocks produce context features $h_C$. The local and context heads predict one envelope and ten Mel bands as $\hat y_L$ and $\hat y_C$. Solid arrows show data flow; dashed arrows show gate conditioning by $h_L$ and $h_C$. Input and output glyphs are schematic.}
\label{fig:architecture}
\end{figure*}

\section{Introduction}
How natural speech is represented in the brain has long been a question in cognitive neuroscience. Over the past two decades, studies have revealed cortical tracking of the speech envelope: auditory cortical activity follows amplitude modulation in the acoustic signal \cite{ahissar2001,luo2007,ding2012}. This response provides a basis for estimating speech acoustics from neural signals. More recently, machine learning has advanced reconstruction of the speech envelope and Mel spectrogram from noninvasive magnetoencephalography (MEG) \cite{wang2026hierarchical} and electroencephalography (EEG) \cite{xu2024conv,puffay2023}.

Speech-envelope reconstruction has used convolutional, Transformer, and dilated-convolution models, including VLAAI, HappyQuokka, and Sea-Wave \cite{accou2023,piao2023,yang2024seawave}. For frequency-resolved Mel-spectrogram reconstruction \cite{bollens2024challenge}, researchers have used convolutional aggregation, state-space modeling, and multiscale fusion to predict band trajectories \cite{xu2024conv,fan2025ssm,fan2025dmf}. Joint envelope--Mel learning and coarse-to-fine acoustic prediction have also been explored \cite{xu2024conv,fang2024}. Hierarchical decoding further supports integrating distinct neural representations \cite{wang2026hierarchical}. A key question is how one reconstructor can retain neural representations distributed across sensors and temporal scales while using context dynamically for each acoustic target.

We address this question with SHINE. A dataset-specific sensor adapter handles differences among EEG and MEG recording arrays. The model then retains intermediate states from different temporal processing depths and learns the contribution of context separately for every acoustic output and time point.

Figure~\ref{fig:architecture} shows the architecture. The residual sensor adapter maps each input array to a common hidden width. Eight dilated temporal blocks progressively expand the receptive field, while the adapter output and four intermediate block states form local features $h_L$. Two pooled-attention context blocks then extract $h_C$. Separate heads produce local hierarchical and context-enhanced estimates, and a learned gate combines them for the envelope and each of ten Mel bands at every time point.

The main contributions are as follows:
\begin{itemize}
\item We develop a sensor-adaptive temporal hierarchy for EEG and MEG. It retains intermediate dilated-block states and fuses local hierarchical and attention-enhanced predictions by acoustic target and time point in joint envelope and ten-band Mel regression.
\item Single-component ablations isolate the effects of the residual sensor adapter and block-internal temporal gate; the adapter shows its clearest controlled gains on the two MEG datasets.
\item Under common data preparation and evaluation, SHINE has the highest mean envelope and mean-Mel correlations among the reported local comparisons on two EEG and two MEG datasets, reaching $r=.5321$ and $r=.1822$ on LibriBrain. It also placed second in the speech-detection Extended Track of the NeurIPS 2025 PNPL Competition \cite{landau2025,pnpl2025results}.
\end{itemize}
\section{Related Work}
Reconstructing the continuous speech envelope is a basic task in speech-feature decoding. Early backward stimulus models estimated the acoustic envelope from EEG to analyze neural tracking of speech \cite{osullivan2015,crosse2016}. Neural models strengthened temporal processing: VLAAI stacks convolutional blocks with skip connections and an output-context layer \cite{accou2023}; HappyQuokka uses a pre-normalized Transformer \cite{piao2023}; Sea-Wave uses dilated convolutions \cite{yang2024seawave}; and DECAF adds an autoregressive prior based on recently reconstructed speech \cite{thakkar2026decaf}. These approaches improve continuous-envelope prediction through different temporal mechanisms. Related auditory-attention decoders such as DARNet and DBPNet target discriminative decisions \cite{yan2024darnet,fan2024dbp}.

The envelope captures broadband amplitude variation, while the Mel spectrogram describes frequency-resolved trajectories \cite{bollens2024challenge}. ConvConcatNet jointly trains envelope and ten-band Mel predictions \cite{xu2024conv}; Cross-Attention-Guided WaveNet predicts envelope and spectral targets in coarse-to-fine stages \cite{fang2024}. SSM2Mel uses state-space layers for long sequences \cite{fan2025ssm}, DMF2Mel fuses temporal scales \cite{fan2025dmf}, and DA2Mel and HFFN-ID introduce subject-adaptive or subject-conditioned processing \cite{fan2026da2mel,huang2026hffn}. SHINE retains several neural temporal depths in one reconstructor and dynamically allocates local and long-range context to the envelope and individual Mel bands.

SHINE combines temporal hierarchy and target-specific context use. A residual sensor adapter maps each dataset's input to a common hidden width; intermediate dilated-block outputs form local hierarchical features; and a target- and time-dependent gate integrates local and pooled-context predictions.

\section{SHINE}
\subsection{Reconstruction objective}
At 64 Hz, SHINE takes a 30-s neural window $x\in\mathbb{R}^{C\times T}$, where $C$ is the dataset-specific sensor count and $T=1920$. It predicts a time-aligned speech envelope and ten Mel bands. As shown in Fig.~\ref{fig:architecture}, a sensor adapter feeds the temporal hierarchy; local and context heads yield separate estimates; and a target-wise gate fuses them. Both branch predictions receive supervision alongside the fused output.

\Needspace{9\baselineskip}
For fused prediction $\hat y=(\hat e,\hat m)$ and target $y=(e,m)$, Pearson correlation is computed per output channel within each window and then averaged over the batch. Auxiliary losses supervise the local and context heads:
\begin{align}
\ell(\hat y,y)&=(1-r_e)+\frac{1}{10}\sum_{b=1}^{10}(1-r_{m,b}),\\
\mathcal L&=\ell(\hat y,y)+0.1\ell(\hat y_L,y)+0.1\ell(\hat y_C,y).
\end{align}
Here $r_e$ is the envelope correlation and $r_{m,b}$ is the correlation of Mel band $b$ for $b=1,\ldots,10$.

\subsection{Residual sensor adapter}
The adapter maps each dataset-specific sensor array to a fixed hidden width:
\begin{equation}
h_0=\operatorname{LN}\!\left[\bigl(\operatorname{Conv}_{1}(x)+f_{\rm main}(x)\bigr)/\sqrt{2}\right],
\end{equation}
Here $h_0$ is the 96-channel adapter output, $\operatorname{Conv}_{1}$ is the shortcut $1\!\times\!1$ projection, and $f_{\rm main}$ contains two $1\!\times\!1$ convolutions separated by GELU. LayerNorm acts over the 96 channels at each time point. The shortcut preserves a direct projection while the main path learns nonlinear sensor mixing. Datasets use the same structure, with a channel-matched first projection and independently trained weights. The full model has 671,745 parameters for 204 input channels, 631,425 for 64 channels, and 629,409 for 57 channels.

\subsection{Temporal hierarchy}
Let $h_i$ denote the output of temporal block $i=1,\ldots,8$. Each 96-channel block applies a width-9 depthwise convolution with dilation $d_i=2^{i-1}$, giving rates 1 through 128. The convolution splits into value and gate tensors $v_i$ and $g_{\mathrm{temp},i}$, followed by $\tanh(v_i)\odot\sigma(g_{\mathrm{temp},i})$ and a two-layer $1\!\times\!1$ channel mixer $f_i$. The update is $h_i=\operatorname{LN}[h_{i-1}+0.5f_i\{\tanh(v_i)\odot\sigma(g_{\mathrm{temp},i})\}]$. The theoretical receptive field is 2041 samples, exceeding the 1920-sample window; boundary outputs also use zero-padded positions. We concatenate $h_0,h_2,h_4,h_6,h_8$ and apply a $1\!\times\!1$ projection, normalization, and GELU to obtain local features $h_L\in\mathbb{R}^{96\times T}$.

\subsection{Local and context prediction}
Two residual context blocks map $h_L$ to $h_C$. Each first adds $0.2$ times a width-31 depthwise positional convolution. Pre-normalized four-head attention then uses all $T$ query positions and factor-8 average-pooled keys/values with ceiling mode. A pre-normalized GELU feed-forward layer expands the hidden width fourfold; dropout is 0.1. Both LayerNorm operations act over channels at each time point. The local and context heads predict $\hat y_L=W_Lh_L$ and $\hat y_C=W_Ch_C$, each with one envelope and ten Mel outputs. A $1\!\times\!1$ convolution $W_g$ maps $[h_L;h_C]$ to 11 target-wise gate channels. Fusion at every target and time point is
\begin{equation}
g_{\mathrm{fuse}}=\sigma(W_g[h_L;h_C]),\qquad \hat y=\hat y_L+g_{\mathrm{fuse}}\odot(\hat y_C-\hat y_L).
\end{equation}

\Needspace{7\baselineskip}
\section{Experiments}
\subsection{Datasets}
We study continuous-speech reconstruction on two EEG and two MEG datasets. All four follow the same preparation pipeline: 30-Hz low-pass filtering, 64-Hz resampling, channel selection, acoustic alignment, recording-wise standardization, and windowing. Section~4.2 details the target extraction and data steps. Table~\ref{tab:data} summarizes the prepared recordings entering the models.

\textbf{SparrKULee.} This corpus records 64-channel EEG from 85 native Dutch/Flemish listeners who heard audiobooks and podcasts for 90--150 min each \cite{sparrkulee2024}. We use 662 prepared recordings from this corpus.

\textbf{PKUEEG.} Twenty-five native Mandarin listeners heard 50 synthesized chapters of \emph{Romance of the Three Kingdoms} across three recording days, providing approximately four hours of speech per listener \cite{pkueeg2026}. EEG was acquired with NeuroScan on the first two days and Neuracle on the third. The prepared data contribute 1,250 recordings with 57 common scalp EEG channels each.

\textbf{LibriBrain.} A single participant listened to more than 50 hours of English LibriVox audiobooks drawn largely from the Sherlock Holmes stories \cite{ozdogan2025}. The source MEG recordings have 306 Neuromag channels; our 70 prepared recordings retain the 204 planar gradiometers as model input.

\textbf{SMN4Lang.} The same 12 native Mandarin listeners heard 60 naturalistic stories in separate MEG and fMRI sessions, with approximately six hours of story material per listener \cite{smn4lang2022}. We use the MEG portion, represented by 720 prepared subject--story recordings and 204 planar gradiometers per recording.

\begin{table}[H]
\centering
\caption{Prepared recordings in the reported experiments. Split counts are train/validation/test; channels are model inputs, with MEG using planar gradiometers.}
\label{tab:data}
\setlength{\tabcolsep}{2.8pt}
\begin{tabular}{@{}lcccc@{}}
\toprule
Dataset & Modality & Recordings & Train/Val/Test & Channels\\
\midrule
SparrKULee~\cite{sparrkulee2024} & EEG & 662 & 263/207/192 & 64\\
PKUEEG~\cite{pkueeg2026} & EEG & 1250 & 578/304/368 & 57\\
LibriBrain~\cite{ozdogan2025} & MEG & 70 & 50/10/10 & 204\\
SMN4Lang~\cite{smn4lang2022} & MEG & 720 & 336/174/210 & 204\\
\bottomrule
\end{tabular}
\end{table}

\subsection{Data preparation and evaluation}
Neural signals are low-pass filtered at 30 Hz and resampled to 64 Hz. We retain all prepared EEG channels (64 for SparrKULee and 57 for PKUEEG). From each 306-channel Neuromag array, MEG retains triplet offsets 1/2, giving 204 planar gradiometers. Precomputed acoustic targets contain one envelope and ten Mel bands at 64 Hz. The envelope sums 28 gammatone-band magnitudes after $|a|^{0.6}$ compression and receives an 8-Hz zero-phase low-pass filter \cite{vanthornhout2018}. The Mel targets use 25-ms Hann windows, a nominal 128-Hz frame rate, power spectra, and ten Slaney-normalized bands from 0 to 5 kHz, resampled to 64 Hz without log compression.

The resulting recording counts do not follow an exact 70/15/15 split. We independently sort subject and stimulus IDs by SHA-256 hash. For each entity type, about 15\% enters test, 15\% validation, and the rest training, with each held-out count set to $\max(1,\operatorname{round}(.15N))$. When fewer than four subjects are present, all subject IDs are training-assigned and stimulus IDs determine held-out recordings. A recording enters test if either ID is test, otherwise validation if either is validation, and training only if both are training-assigned. Hence no training recording contains a subject or stimulus assigned to validation or test. LibriBrain has one subject, so cross-subject generalization is not evaluated there.

Neural, envelope, and Mel arrays are truncated to their common minimum length. Training windows have length 1920 and stride 960 (50\% overlap); validation and test use length and stride 1920. Before windowing, each array channel is standardized independently over the complete recording's time axis.

At evaluation, all windows and time samples in a split are pooled before one Pearson correlation is computed per output channel; mean-Mel is the arithmetic mean of the ten Mel-band correlations.

\subsection{Training and implementation}
The reported SHINE and component runs use AdamW (learning rate $3\times10^{-4}$, weight decay $10^{-2}$), batch size 8, and 60 epochs for each of seeds 0, 1, and 2. ReduceLROnPlateau uses factor 0.5 and patience 2; the validation sum of envelope and mean-Mel correlations drives the schedule and selects the best checkpoint. CUDA training uses automatic mixed precision and clips the global gradient norm to 5. Python, NumPy, and all CUDA devices are seeded. We implement the networks in PyTorch and train on an NVIDIA A100 GPU.

All component variants share the same data, objective, 60-epoch schedule, and checkpoint rule as SHINE; each dataset has independently trained weights.

\subsection{Comparison implementations}
The nine local comparison implementations share data splits, acoustic targets, windows, and metrics with SHINE. They are BrainMagick \cite{defossez2023}, VLAAI \cite{accou2023}, HappyQuokka \cite{piao2023}, ConvConcatNet \cite{xu2024conv}, SSM2Mel \cite{fan2025ssm}, DMF2Mel \cite{fan2025dmf}, AWaveNet \cite{yang2024seawave,fang2024}, CNN--LSTM \cite{fu2021crnn}, and DilatedConv \cite{accou2020dilated}. The first three are extended to 204-channel MEG; ConvConcatNet, SSM2Mel, and DMF2Mel are adapted to joint envelope--Mel targets. AWaveNet, CNN--LSTM, and DilatedConv are internal controls based on the cited architectures, not official implementations. All nine train for 60 epochs with validation-based checkpoint selection; SSM2Mel and DMF2Mel retain model-specific batch sizes and learning rates.

\Needspace{6\baselineskip}
\section{Results and Discussion}
Tables~\ref{tab:main} and~\ref{tab:ablation} report the archived test measurements as means and sample standard deviations over seeds 0, 1, and 2 under the training schedules described in Section~4.

\subsection{Main results}
SHINE has the highest three-seed mean among the nine local baselines in all eight metric--dataset cells of Table~\ref{tab:main}. For SparrKULee, PKUEEG, LibriBrain, and SMN4Lang, respectively, its envelope margins over the strongest baseline means are $.0177$, $.0167$, $.0063$, and $.0270$; the corresponding mean-Mel margins are $.0028$, $.0063$, $.0062$, and $.0077$. This coverage spans both EEG and MEG and both envelope and spectral-trajectory reconstruction.

\begin{table*}[!t]
\centering
\caption{Main comparison: pooled test Pearson $r$ (mean $\pm$ sample standard deviation over seeds 0/1/2). Mean Mel is the average over ten band-wise correlations. Comparison rows are local implementations on the common data and metric. Bold marks SHINE means; underlining marks the strongest comparison means.}
\label{tab:main}
\setlength{\tabcolsep}{1.5pt}
\let\standardpm\pm
\renewcommand{\pm}{\mathord{\standardpm}}
\begin{tabular}{@{}l*{8}{c}@{}}
\toprule
& \multicolumn{4}{c}{Speech envelope (Pearson $r$)} & \multicolumn{4}{c}{Mean Mel-spectrogram (Pearson $r$)}\\
\cmidrule(lr){2-5}\cmidrule(lr){6-9}
Model & \shortstack{Sparr\\KULee} & \shortstack{PKU\\EEG} & \shortstack{Libri\\Brain} & \shortstack{SMN4\\Lang} & \shortstack{Sparr\\KULee} & \shortstack{PKU\\EEG} & \shortstack{Libri\\Brain} & \shortstack{SMN4\\Lang}\\
\midrule
SHINE & $\mathbf{.1877}\pm.0005$ & $\mathbf{.2300}\pm.0093$ & $\mathbf{.5321}\pm.0046$ & $\mathbf{.3594}\pm.0074$ & $\mathbf{.0483}\pm.0001$ & $\mathbf{.0871}\pm.0036$ & $\mathbf{.1822}\pm.0016$ & $\mathbf{.1370}\pm.0034$\\
\midrule
BrainMagick & $.1698\pm.0009$ & $.1868\pm.0082$ & $.4770\pm.0057$ & $.2772\pm.0033$ & $\underline{.0455}\pm.0003$ & $.0722\pm.0031$ & $.1608\pm.0029$ & $.1113\pm.0018$\\
VLAAI & $.1581\pm.0057$ & $\underline{.2133}\pm.0019$ & $.5162\pm.0005$ & $\underline{.3324}\pm.0010$ & $.0424\pm.0009$ & $\underline{.0808}\pm.0009$ & $.1742\pm.0004$ & $\underline{.1293}\pm.0013$\\
HappyQuokka & $.1524\pm.0047$ & $.1963\pm.0033$ & $\underline{.5258}\pm.0007$ & $.3111\pm.0084$ & $.0403\pm.0010$ & $.0737\pm.0022$ & $\underline{.1760}\pm.0005$ & $.1166\pm.0037$\\
ConvConcatNet & $.1590\pm.0067$ & $.1930\pm.0026$ & $.4475\pm.0070$ & $.2062\pm.0101$ & $.0427\pm.0015$ & $.0755\pm.0003$ & $.1479\pm.0026$ & $.0869\pm.0032$\\
AWaveNet & $\underline{.1700}\pm.0017$ & $.2039\pm.0077$ & $.4360\pm.0058$ & $.0288\pm.0002$ & $.0446\pm.0002$ & $.0767\pm.0041$ & $.1437\pm.0026$ & $.0137\pm.0003$\\
SSM2Mel & $.1582\pm.0014$ & $.1457\pm.0105$ & $.4691\pm.0033$ & $.1535\pm.1163$ & $.0417\pm.0008$ & $.0560\pm.0036$ & $.1540\pm.0018$ & $.0624\pm.0483$\\
DMF2Mel & $.1523\pm.0031$ & $.1630\pm.0125$ & $.4632\pm.0055$ & $.2225\pm.0119$ & $.0407\pm.0009$ & $.0628\pm.0046$ & $.1518\pm.0017$ & $.0908\pm.0040$\\
CNN--LSTM & $.1144\pm.0012$ & $.1171\pm.0047$ & $.3183\pm.0024$ & $.0665\pm.0020$ & $.0314\pm.0005$ & $.0387\pm.0011$ & $.1064\pm.0014$ & $.0286\pm.0018$\\
DilatedConv & $.1178\pm.0002$ & $.1286\pm.0044$ & $.3091\pm.0070$ & $.0648\pm.0492$ & $.0353\pm.0002$ & $.0518\pm.0014$ & $.1048\pm.0023$ & $.0248\pm.0209$\\
\bottomrule
\end{tabular}
\let\pm\standardpm
\end{table*}

\begin{table*}[!t]
\centering
\caption{Component analysis at 60 epochs: pooled test Pearson $r$ (mean $\pm$ sample standard deviation over seeds 0/1/2). Mean Mel averages ten band-wise correlations. Bold marks SHINE means; underlining marks the strongest variant means.}
\label{tab:ablation}
\setlength{\tabcolsep}{1.0pt}
\let\standardpm\pm
\renewcommand{\pm}{\mathord{\standardpm}}
\begin{tabular}{@{}l*{8}{c}@{}}
\toprule
& \multicolumn{4}{c}{Speech envelope (Pearson $r$)} & \multicolumn{4}{c}{Mean Mel-spectrogram (Pearson $r$)}\\
\cmidrule(lr){2-5}\cmidrule(lr){6-9}
Variant & \shortstack{Sparr\\KULee} & \shortstack{PKU\\EEG} & \shortstack{Libri\\Brain} & \shortstack{SMN4\\Lang} & \shortstack{Sparr\\KULee} & \shortstack{PKU\\EEG} & \shortstack{Libri\\Brain} & \shortstack{SMN4\\Lang}\\
\midrule
SHINE & $\mathbf{.1877}\pm.0005$ & $\mathbf{.2300}\pm.0093$ & $\mathbf{.5321}\pm.0046$ & $\mathbf{.3594}\pm.0074$ & $\mathbf{.0483}\pm.0001$ & $\mathbf{.0871}\pm.0036$ & $\mathbf{.1822}\pm.0016$ & $\mathbf{.1370}\pm.0034$\\
\midrule
\shortstack[l]{Single spatial\\projection} & $\underline{.1876}\pm.0024$ & $.2264\pm.0051$ & $.5128\pm.0057$ & $.3159\pm.0019$ & $\underline{.0482}\pm.0003$ & $.0849\pm.0032$ & $.1752\pm.0025$ & $.1243\pm.0011$\\
No block gate & $.1809\pm.0019$ & $\underline{.2285}\pm.0007$ & $\underline{.5264}\pm.0047$ & $\underline{.3591}\pm.0042$ & $.0473\pm.0006$ & $\underline{.0870}\pm.0007$ & $\underline{.1807}\pm.0017$ & $\underline{.1369}\pm.0010$\\
\bottomrule
\end{tabular}
\end{table*}

\Needspace{8\baselineskip}
\subsection{Component analysis}
Table~\ref{tab:ablation} compares two single-change variants under the same training protocol. Single spatial projection replaces the residual adapter with one $1\!\times\!1$ convolution, isolating nonlinear residual sensor mixing while retaining the temporal hierarchy and two-head fusion. No block gate removes the block-internal $\sigma(g_{\mathrm{temp}})$ but retains $\tanh(v)$ and the final target--time fusion gate $g_{\mathrm{fuse}}$. The adapter shows its clearest controlled gains on MEG: against a single spatial projection, SHINE increases envelope/mean-Mel correlation by $.0193/.0070$ on LibriBrain and $.0435/.0127$ on SMN4Lang.

The block-internal temporal gate raises the mean envelope correlation on all four datasets, most notably by $.0068$ on SparrKULee and $.0057$ on LibriBrain. Thus the component comparison identifies sensor adaptation as the clearest contributor to the cross-modal architecture, especially for MEG.

\Needspace{5\baselineskip}
\subsection{Limitations}
LibriBrain includes one subject, so cross-subject generalization cannot be assessed on that dataset. Standard deviations across three seeds characterize variation across training runs but do not estimate uncertainty across independent recordings or subjects.

\FloatBarrier
\section{Conclusion}
We presented SHINE for reconstructing the speech envelope and ten Mel bands from EEG and MEG. Its sensor adapter handles different recording arrays, intermediate states retain successive temporal depths, and target- and time-dependent fusion combines local hierarchical and attention-enhanced context estimates. Across two EEG and two MEG datasets, SHINE has the highest mean envelope and mean-Mel Pearson correlations among the nine local baseline implementations; the residual adapter provides the clearest controlled gains on MEG. Future work will examine competing talkers, noise, reverberation, and reconstruction across sessions and subjects.

\section{Acknowledgement}
This work was supported by the STI 2030---Major Projects (No. 2021ZD0201500), the High-performance Computing Platform of Peking University and the Biomedical Computing Platform of National Biomedical Imaging Center, Peking University.

\bibliographystyle{IEEEbib}
\bibliography{shine_refs}

\begin{thebibliography}{10}

\bibitem{ahissar2001}
E.~Ahissar, S.~Nagarajan, M.~Ahissar, A.~Protopapas, H.~Mahncke, and M.~M.
  Merzenich,
\newblock ``Speech comprehension is correlated with temporal response patterns
  recorded from auditory cortex,''
\newblock {\em Proc. Natl. Acad. Sci. USA}, vol. 98, no. 23, pp. 13367--13372,
  2001.

\bibitem{luo2007}
H.~Luo and D.~Poeppel,
\newblock ``Phase patterns of neuronal responses reliably discriminate speech
  in human auditory cortex,''
\newblock {\em Neuron}, vol. 54, no. 6, pp. 1001--1010, 2007.

\bibitem{ding2012}
N.~Ding and J.~Z. Simon,
\newblock ``Neural coding of continuous speech in auditory cortex during
  monaural and dichotic listening,''
\newblock {\em Journal of Neurophysiology}, vol. 107, no. 1, pp. 78--89, 2012.

\bibitem{wang2026hierarchical}
B.~Wang, X.~Xu, B.~Xiao, L.~Zheng, X.~Wu, H.~Cheng, and J.~Chen,
\newblock ``Hierarchical decoding of perceived speech from non-invasive brain
  recordings,''
\newblock {\em IEEE Trans. Neural Syst. Rehabil. Eng.}, vol. 34, pp.
  2349--2360, 2026.

\bibitem{xu2024conv}
X.~Xu, B.~Wang, Y.~Yan, H.~Zhu, Z.~Zhang, X.~Wu, and J.~Chen,
\newblock ``{ConvConcatNet}: A deep convolutional neural network to reconstruct
  {Mel} spectrogram from the {EEG},''
\newblock in {\em Proc. IEEE Int. Conf. Acoust., Speech Signal Process.
  Workshops ({ICASSPW})}, 2024, pp. 113--114.

\bibitem{puffay2023}
C.~Puffay, B.~Accou, L.~Bollens, M.~Jalilpour Monesi, J.~Vanthornhout, H.~Van
  hamme, and T.~Francart,
\newblock ``Relating {EEG} to continuous speech using deep neural networks: a
  review,''
\newblock {\em Journal of Neural Engineering}, vol. 20, no. 4, pp. 041003,
  2023.

\bibitem{accou2023}
B.~Accou, J.~Vanthornhout, H.~Van hamme, and T.~Francart,
\newblock ``Decoding of the speech envelope from {EEG} using the {VLAAI} deep
  neural network,''
\newblock {\em Scientific Reports}, vol. 13, no. 1, pp. 812, 2023.

\bibitem{piao2023}
Z.~Piao, M.~Kim, H.~Yoon, and H.-G. Kang,
\newblock ``{HappyQuokka} system for {ICASSP} 2023 auditory {EEG} challenge,''
\newblock in {\em Proc. IEEE Int. Conf. Acoust., Speech Signal Process.
  ({ICASSP})}, 2023, pp. 1--2.

\bibitem{yang2024seawave}
L.~Yang, B.~Van Dyck, and M.~M.~Van Hulle,
\newblock ``{Sea-Wave}: Speech envelope reconstruction from auditory {EEG} with
  an adapted {WaveNet},''
\newblock {\em {IEEE} Open Journal of Signal Processing}, vol. 5, pp. 686--699,
  2024.

\bibitem{bollens2024challenge}
L.~Bollens, C.~Puffay, B.~Accou, J.~Vanthornhout, H.~Van hamme, and
  T.~Francart,
\newblock ``{ICASSP} 2024 auditory {EEG} decoding challenge,''
\newblock in {\em Proc. IEEE Int. Conf. Acoust., Speech Signal Process.
  Workshops ({ICASSPW})}, 2024, pp. 127--128.

\bibitem{fan2025ssm}
C.~Fan, S.~Zhang, J.~Zhang, Z.~Pan, and Z.~Lv,
\newblock ``{SSM2Mel}: State space model to reconstruct {Mel} spectrogram from
  the {EEG},''
\newblock in {\em Proc. IEEE Int. Conf. Acoust., Speech Signal Process.
  ({ICASSP})}, 2025, pp. 1--5.

\bibitem{fan2025dmf}
C.~Fan, S.~Zhang, J.~Zhang, E.~Liu, X.~Li, G.~Zhao, and Z.~Lv,
\newblock ``{DMF2Mel}: A dynamic multiscale fusion network for {EEG}-driven
  {Mel} spectrogram reconstruction,''
\newblock in {\em Proc. ACM Int. Conf. Multimedia ({MM})}, 2025, pp.
  6977--6985.

\bibitem{fang2024}
Y.~Fang, H.~Li, X.~Zhang, F.~Chen, and G.~Gao,
\newblock ``Cross-attention-guided {WaveNet} for {Mel} spectrogram
  reconstruction in the {ICASSP} 2024 auditory {EEG} challenge,''
\newblock in {\em Proc. IEEE Int. Conf. Acoust., Speech Signal Process.
  Workshops ({ICASSPW})}, 2024, pp. 7--8.

\bibitem{landau2025}
G.~Landau, M.~{\"O}zdogan, G.~Elvers, F.~Mantegna, P.~Somaiya, D.~Jayalath,
  et~al.,
\newblock ``The 2025 {PNPL} competition: Speech detection and phoneme
  classification in the {LibriBrain} dataset,''
\newblock in {\em NeurIPS 2025 Competition Track}, 2025,
\newblock \href{https://arxiv.org/abs/2506.10165}{arXiv:2506.10165}.

\bibitem{pnpl2025results}
{Parker Jones Neural Processing Lab},
\newblock ``{2025 PNPL Competition Winners},'' Official results,
  \href{https://neural-processing-lab.github.io/2025-libribrain-competition/editions/2025/prizes/}{online},
  2025.

\bibitem{osullivan2015}
J.~A. O'Sullivan, A.~J. Power, N.~Mesgarani, S.~Rajaram, J.~J. Foxe, B.~G.
  Shinn-Cunningham, M.~Slaney, S.~A. Shamma, and E.~C. Lalor,
\newblock ``Attentional selection in a cocktail party environment can be
  decoded from single-trial {EEG},''
\newblock {\em Cerebral Cortex}, vol. 25, no. 7, pp. 1697--1706, 2015.

\bibitem{crosse2016}
M.~J. Crosse, G.~M.~Di Liberto, A.~Bednar, and E.~C. Lalor,
\newblock ``The multivariate temporal response function ({mTRF}) toolbox: A
  {MATLAB} toolbox for relating neural signals to continuous stimuli,''
\newblock {\em Frontiers in Human Neuroscience}, vol. 10, pp. 604, 2016.

\bibitem{thakkar2026decaf}
K.~Thakkar and M.~Elhilali,
\newblock ``{DECAF}: Dynamic envelope context-aware fusion for speech-envelope
  reconstruction from {EEG},''
\newblock in {\em Proc. IEEE Int. Conf. Acoust., Speech Signal Process.
  ({ICASSP})}, 2026, pp. 1--5.

\bibitem{yan2024darnet}
S.~Yan, C.~Fan, H.~Zhang, X.~Yang, J.~Tao, and Z.~Lv,
\newblock ``{DARNet}: Dual attention refinement network with spatiotemporal
  construction for auditory attention detection,''
\newblock in {\em Advances in Neural Information Processing Systems}, 2024,
  vol.~37, pp. 31688--31707.

\bibitem{fan2024dbp}
Q.~Ni, H.~Zhang, C.~Fan, S.~Pei, C.~Zhou, and Z.~Lv,
\newblock ``{DBPNet}: Dual-branch parallel network with temporal-frequency
  fusion for auditory attention detection,''
\newblock in {\em Proceedings of the Thirty-Third International Joint
  Conference on Artificial Intelligence ({IJCAI})}, 2024, pp. 3115--3123.

\bibitem{fan2026da2mel}
C.~Fan, H.~Lv, S.~Zhang, J.~Zhou, X.~Li, C.~Sun, and Z.~Lv,
\newblock ``{DA2Mel}: A dual-dynamic adaptive network for {EEG}-to-{Mel}
  spectrogram reconstruction,''
\newblock {\em ACM Trans. Auton. Adapt. Syst.}, 2026.

\bibitem{huang2026hffn}
C.~Huang, Z.~Liu, Y.~Peng, and W.~Kong,
\newblock ``{HFFN-ID}: A hierarchical feature fusion network with bi-phase
  subject {ID} modulation for {EEG} {Mel}-spectrogram reconstruction,''
\newblock in {\em Proc. Int. Joint Conf. Artif. Intell. ({IJCAI})}, 2026, pp.
  3612--3620.

\bibitem{sparrkulee2024}
B.~Accou, L.~Bollens, M.~Gillis, W.~Verheijen, H.~Van hamme, and T.~Francart,
\newblock ``{SparrKULee}: A speech-evoked auditory response repository from {KU
  Leuven}, containing the {EEG} of 85 participants,''
\newblock {\em Data}, vol. 9, no. 8, pp. 94, 2024.

\bibitem{pkueeg2026}
B.~Xiao, X.~Xu, Y.~Yan, S.~Li, Z.~Zhang, L.~Zheng, J.~Liang, X.~Wu, H.~Cheng,
  and J.~Chen,
\newblock ``{PKUEEG},'' OpenNeuro, version 1.0.3 [dataset], 2026,
\newblock
  \href{https://doi.org/10.18112/openneuro.ds008834.v1.0.3}{doi:10.18112/openneuro.ds008834.v1.0.3}.

\bibitem{ozdogan2025}
M.~{\"O}zdogan, G.~Landau, G.~Elvers, D.~Jayalath, P.~Somaiya, F.~Mantegna,
  M.~Woolrich, and O.~Parker~Jones,
\newblock ``{LibriBrain}: Over 50 hours of within-subject {MEG} to improve
  speech decoding methods at scale,''
\newblock in {\em Proc. NeurIPS Datasets and Benchmarks Track}, 2025, vol.~38,
  pp. 17016--17053,
\newblock
  \href{https://doi.org/10.52202/085713-0511}{doi:10.52202/085713-0511}.

\bibitem{smn4lang2022}
S.~Wang, X.~Zhang, J.~Zhang, and C.~Zong,
\newblock ``A synchronized multimodal neuroimaging dataset for studying brain
  language processing,''
\newblock {\em Scientific Data}, vol. 9, no. 1, pp. 590, 2022.

\bibitem{vanthornhout2018}
J.~Vanthornhout, L.~Decruy, J.~Wouters, J.~Z. Simon, and T.~Francart,
\newblock ``Speech intelligibility predicted from neural entrainment of the
  speech envelope,''
\newblock {\em J. Assoc. Res. Otolaryngol.}, vol. 19, no. 2, pp. 181--191,
  2018.

\bibitem{defossez2023}
A.~D{\'e}fossez, C.~Caucheteux, J.~Rapin, O.~Kabeli, and J.-R. King,
\newblock ``Decoding speech perception from non-invasive brain recordings,''
\newblock {\em Nature Machine Intelligence}, vol. 5, no. 10, pp. 1097--1107,
  2023.

\bibitem{fu2021crnn}
Z.~Fu, B.~Wang, X.~Wu, and J.~Chen,
\newblock ``Auditory attention decoding from {EEG} using convolutional
  recurrent neural network,''
\newblock in {\em Proc. 29th Eur. Signal Process. Conf. ({EUSIPCO})}, 2021, pp.
  970--974.

\bibitem{accou2020dilated}
B.~Accou, M.~Jalilpour Monesi, J.~Montoya, H.~Van hamme, and T.~Francart,
\newblock ``Modeling the relationship between acoustic stimulus and {EEG} with
  a dilated convolutional neural network,''
\newblock in {\em Proc. 28th Eur. Signal Process. Conf. ({EUSIPCO} 2020)},
  2021, pp. 1175--1179.

\end{thebibliography}
\end{document}